\documentclass[11pt]{article}
\usepackage{csquotes}
\usepackage{times}
\usepackage{geometry}
\usepackage{enumitem,amssymb}
\usepackage{ragged2e}
\newlist{thematic}{itemize}{8}
\setlist[thematic]{label=$\square$}
\usepackage{pifont}
\usepackage{setspace}
\usepackage{eso-pic,graphicx}
\usepackage{tikz}
\usepackage{atbegshi}
\usepackage{url}
\usepackage{xcolor}    
\usepackage{multicol}
\usepackage[english]{babel}
\usepackage{enumitem}
\usepackage{changepage}

\usepackage{titlesec}

\titlespacing*{\section}
  {0pt}   % left indent
  {1.5ex plus 0.5ex minus 0.5ex}   % space before
  {0.8ex plus 0.3ex minus 0.3ex}   % space after

\usepackage[backend=biber,
    style=numeric,
    sorting=none,
    maxnames=1,   
    minnames=1]{biblatex} 
\renewbibmacro{in:}{,}

\AtEveryBibitem{%
  \clearfield{title}%
  \clearfield{subtitle}%
  \clearfield{doi}%
  \clearfield{url}
  \clearfield{eprintclass}
}

\definecolor{myblue}{RGB}{0,121,194}
\usepackage[
    colorlinks=true,  
    citecolor=myblue,
    linkcolor=myblue,      
    urlcolor=myblue        
]{hyperref}

\renewbibmacro*{url+urldate}{%
  \iffieldundef{url}{}{%
    \printfield{url}%
  }%
}

\AddToShipoutPictureBG{%
  \ifnum\value{page}=1
    \AtPageLowerLeft{%
      \includegraphics[width=\paperwidth]{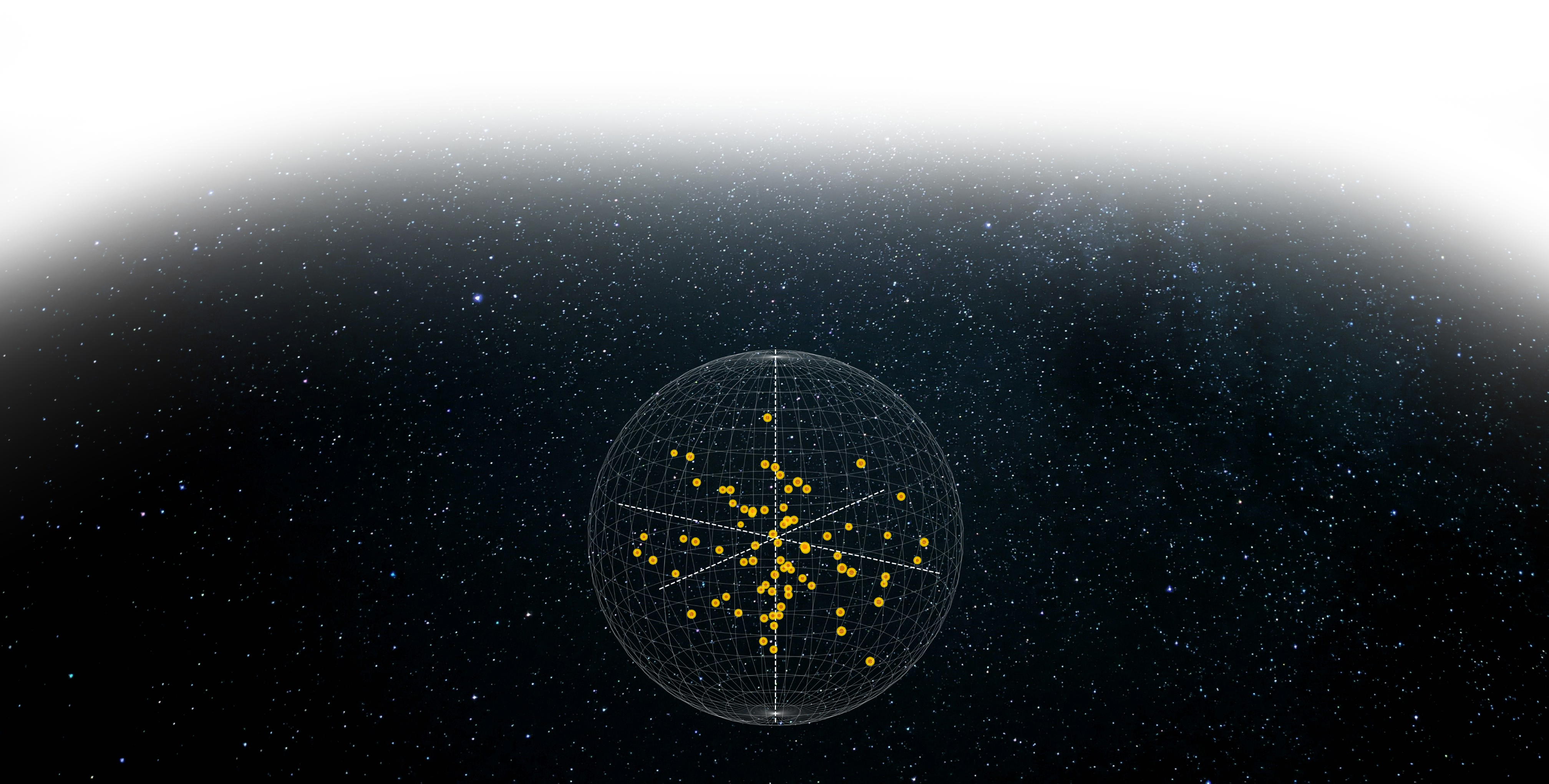}%
    }%
  \fi
}

\begin{document}

\selectlanguage{english}
\begin{adjustwidth}{1.5cm}{1.5cm}
{
\raggedright
\Large
ESO Expanding Horizons initiative 2025 \\
Call for White Papers

\vspace{2.cm}
\begin{spacing}{1.6}
\textbf{\fontsize{22pt}{40pt}\selectfont
Beyond general relativity: probing gravity with gravitational redshifts}
\end{spacing}
\normalsize
\vspace{0.5cm}
\textbf{Authors:} E. Tsaprazi$^\mathbf{1}$, G. F. Lesci$^\mathbf{2,3}$, F. Marulli$^\mathbf{2,3,4}$, A. F. Heavens$^\mathbf{1}$, P. Rosati$^\mathbf{5,3}$, S. Contarini$^\mathbf{6}$, E. A. Maraboli$^\mathbf{7}$, P. Dayal$^\mathbf{8}$, O. Lahav$^\mathbf{9}$, L. Moscardini$^\mathbf{2}$

\vspace{0.3cm}
\textbf{Contact:} \href{mailto:e.tsaprazi@imperial.ac.uk}{e.tsaprazi@imperial.ac.uk} 
\linebreak 

\textbf{Affiliations:} \\
$^\mathbf{1}$ Imperial Centre for Inference and Cosmology (ICIC) \& Astrophysics group, Department of Physics, Imperial College, Blackett Laboratory, Prince Consort Road, London SW7 2AZ, UK \\
$^\mathbf{2}$ Dipartimento di Fisica e Astronomia “Augusto Righi” - Alma Mater Studiorum Università di Bologna, via Piero Gobetti 93/2, I-40129 Bologna, Italy \\
$^\mathbf{3}$ INAF - Osservatorio di Astrofisica e Scienza dello Spazio di Bologna, via Piero Gobetti 93/3, I-40129 Bologna, Italy\\
$^\mathbf{4}$ INFN - Sezione di Bologna, viale Berti Pichat 6/2, I-40127 Bologna, Italy \\
$^\mathbf{5}$ Department of Physics and Earth Sciences, University of Ferrara, Via G. Saragat, 1, 44122 Ferrara, Italy\\
$^\mathbf{6}$ Max-Planck-Institut für extraterrestrische Physik, Postfach 1312, Giessenbachstr., 85748 Garching, Germany\\
$^\mathbf{7}$ Dipartimento di Fisica, Università degli Studi di Milano, Via Celoria 16, I-20133 Milano, Italy\\
$^\mathbf{8}$ Canadian Institute for Theoretical Astrophysics, 60 St George St, University\\
$^\mathbf{9}$ Department of Physics and Astronomy, University College London,
Gower Street, London WC1E 6BT, UK\\}
\end{adjustwidth}
\pagenumbering{gobble} 

\pagebreak

\section*{Introduction}
General relativity (GR) has stood as the foundational theory of gravity, accurately describing phenomena from planetary orbits to gravitational lensing. While GR has been remarkably successful in explaining a wide range of observations, cosmological tests of gravity remain limited by instrumental precision and theoretical modelling. In light of persistent tensions and outstanding open questions in our concordance cosmological model, it is crucial to explore whether modified theories of gravity could be favoured over GR when confronted with cosmological tests. Evidence suggests that modifications to GR could alleviate the Hubble tension and offer insights into the nature of dark matter and dark energy \cite{1,2}. If substantiated, this would constitute a profound shift in our understanding of physics. However, the robustness of such results remains under active scrutiny. It is therefore necessary to exploit current cosmological probes optimally and orthogonally to their existing uses.  

Galaxy clusters represent the largest gravitationally bound structures known in the Universe. Their properties are sensitive to structure formation, which is governed by the theory of gravity. Traditionally, galaxy clusters have been used to constrain modified gravity through their abundance and clustering \cite{3}, but these probes are limited by our ability to model nonlinear structure growth and redshift-space distortions (RSD). With the advent of spectroscopic Stage-V surveys, we will be able to precisely probe modified gravity on megaparsec scales using the gravitational redshifts of cluster member galaxies (around 10 km $s^{-1}$), partly alleviating the need for precise small-scale RSD modelling, while introducing a different set of systematic effects tied to baryonic and dynamical physics. As photons propagate in the gravitational potential of clusters, their wavelength is governed by the strength of the potential. The effect of modified gravity can, for certain theories, be modelled as a re-scaling of the gravitational constant, G, by the modified gravity parameter, $\alpha$, leaving an imprint on the redshifts of cluster member galaxies. This approach, therefore, opens a direct avenue to test gravity on small cosmological scales, which can be complementary to more traditional large-scale structure analyses \cite{4,5}.

The dominant driver of uncertainties for this probe is the number of cluster members identified and the precision and accuracy of the cluster-centre determination. The effect is too weak to measure from single clusters, but can be recovered in large samples of stacked clusters. To ensure a clean measurement, these samples should be carefully selected, typically limited in volume and restricted to a well-defined lower mass threshold, to minimise contamination from outliers and misidentified systems. At the same time, the signal is sensitive to the selection of the cluster centre and the calibration of the cluster mass. This determination typically limits the available number of clusters to those with a sufficient number of galaxy members from which the centre can be computed. The requirement for a sufficiently complete and pure galaxy sample imposes additional mass selections. The presence of these systematic effects significantly reduces the current robustness and constraining power of gravitational redshifts as a probe of modified gravity. 

In the coming decade, we expect to be able to apply this probe to substantially more informative and cleaner data sets, thanks to Stage-V cosmological experiments. Specifically, we require:
\begin{itemize}[itemsep=0pt, topsep=0pt, parsep=0pt]
\item wide spectroscopic redshift surveys for a large number of clusters and galaxies,
\item precise and accurate spectroscopic redshifts of cluster member galaxies, 
\item accurate calibration of cluster masses, and low rates of false cluster identification,
\item accurate determination of cluster centres and control of membership systematic effects.
\end{itemize}

\section*{Observational status}

The current gold standard is a sample of 45 000 clusters with spectroscopically identified bright central galaxies out to z=0.5, obtained by cross-correlating SDSS DR16 spectroscopy with photometric clusters \cite{6}. The Rubin Legacy Survey of Space and Time (LSST) \cite{7}, Euclid \cite{8} and the Dark Energy Spectroscopic Instrument surveys \cite{9} will greatly increase the number of detected clusters by 2030, but will still lack homogeneous, high-completeness spectroscopy for member galaxies \cite{10}, so they are unlikely on their own to deliver order-of-magnitude improvements in gravitational-redshift constraints. As a result, we turn our attention to planned facilities in the 2040s and beyond that will combine large survey areas with multi-object spectrographs (MOS) and, in some cases, integral field spectroscopy (IFS). For cluster gravitational-redshift studies, IFS provides spatially resolved views of cluster cores and outskirts. In practice, IFS yields a datacube for each cluster, from which one can construct 2-dimensional maps of the line-of-sight velocity and velocity-dispersion fields. We investigate whether such data can mitigate systematic effects in gravitational-redshift measurements. Our analysis focuses on two measurements, namely: calibrating mis-centring by exploiting a kinematic definition of the cluster centre from the IFS velocity maps, and adopting a radial density profile that more accurately reflects the underlying cluster structure. We find that IFS follow-up cannot deliver a clean calibration of mis-centring in velocity space: the inferred offsets between the BCG and the true centre remain dominated by peculiar velocities, even for large IFS subsamples. Further, even strongly baryonised inner profiles leave the stacked gravitational-redshift signal statistically consistent with a standard Navarro–Frenk–White profile, implying only marginal gains from refining the radial density profile with IFS observations. These limitations highlight that IFS, while valuable, is not sufficient on its own to overcome the dominant systematic effects; its impact becomes significant only when embedded in a dedicated cluster-centric survey strategy. A possible benefit may arise from targeting a subsample of massive, highly concentrated clusters, though this scenario requires investigation within a more realistic survey framework.

By the 2030s, the all-sky Spectroscopic Stage-5 Experiment (Spec-S5, 2036-) \cite{11} and the MUltiplexed Survey Telescope (MUST, 2030s) \cite{12} are expected to provide large-area spectroscopy at moderate resolution, on top of the photometric cluster catalogues from Euclid and Rubin LSST. In parallel, small-field IFS instruments such as ELT/HARMONI \cite{13}, GMTIFS \cite{14}, TMT/IRIS \cite{15} and VLT/BlueMUSE \cite{16} could deliver detailed spectroscopy for a limited number of clusters. In Figure 1a-b, we present a comparison between the constraints expected on modified gravity theories from data sets assuming spectroscopic and photometric redshift uncertainties. While the error bars rapidly decrease from the photometric to spectroscopic survey, there exists a lower threshold in the precision we can expect owing to the presence of member galaxy velocity dispersions, yielding broadly similar constraints for Stage-IV- and Stage-V-like spectroscopy. Therefore, facilities in the 2030s can enable the first generation of high-precision stacked gravitational-redshift measurements, provided their survey strategies devote sufficient resources to building spectroscopic cluster catalogues. 

However, state-of-the-art spectroscopic facilities planned for the 2030s are currently fundamentally galaxy-centric: their baseline survey strategies are optimised for baryon acoustic oscillations, RSD and general large-scale structure, with cluster samples and memberships reconstructed ex post rather than observed through a dedicated, homogeneous cluster programme. In the 2040s, the Wide-Field Spectroscopic Telescope (WST) \cite{17} is instead conceived with an explicit cluster-centric component, in which the high-multiplex MOS and wide-field IFS are deployed to map massive clusters and their environments over a wide redshift range as a core science driver rather than a by-product. The resulting cluster catalogue will be built from the spectroscopic follow-up of clusters identified in LSST. This allows the survey strategy, fibre allocation and depth to be optimised explicitly for uniform, high-completeness sampling of cluster cores and outskirts across redshift and mass, rather than relying on whatever incomplete and inhomogeneous coverage incidentally results from galaxy-centric targeting. By contrast, galaxy-centric surveys optimise fibre usage for a roughly uniform galaxy density on the sky, so clusters receive only whatever spectroscopic sampling happens to arise from those priorities, with strongly variable member counts, radial coverage and purity from halo to halo.

\section*{Technology requirements}

Building on this, we propose that facilities with dedicated cluster programmes, which are currently only planned by WST in the 2040s, include a high-purity spectroscopic cluster catalogue. This would require MOS fibres allocated to densely sample members and infalling galaxies out to 2-5 Mpc, supported by a halo sample with beam/point-spread functions small enough that distinct structures along the line of sight are not blended into single objects. Additionally, we require a well-calibrated selection function with false-positive rate at the few-percent level. In Figure 1c, we compare three completeness scenarios up to redshift z = 2: a catalogue with 100\% spectroscopic completeness, a galaxy-survey–like MOS setup in which clusters are assembled from a generic galaxy redshift survey with radially varying completeness, and a cluster-survey–like MOS setup mimicking a dedicated high-purity cluster programme with very high completeness in the core and elevated completeness at large radii. For the galaxy-survey setting we assume 90\% completeness which drops to 50\% at 4$r_\mathrm{500}$ and 8\% at 4$r_\mathrm{500}$. In our mock catalogues, the median $r_\mathrm{500}$ is 0.4 Mpc. For the cluster-survey setting we assume 90\% completeness out to 1$r_\mathrm{500}$ and 40\% beyond. While optimistic for a galaxy survey, a targeted cluster survey could achieve a high completeness with enough colour bands. In our setting, this produces an improvement on the stacked signal error bars of 7\% in the centre and 50\% in the outskirts compared to a galaxy-like survey. 

Even in the perfect completeness case, we obtain 9\% precision on $\alpha$, which will worsen as richness cuts are applied. This is partly because the strength of constraints depends on modelling deviations from GR through the linear factor, $\alpha$. As can be seen in Figure 1b, this model is not particularly sensitive to small deviations from GR. Therefore, in parallel to the above technical developments, advancements on the modelling side should also be made to explore different parameterisations \cite{18}. The Strategic Roadmap for the next decade of European Astronomy places the search for deviations from GR among the most transformative questions of the future scientific landscape. The facilities expected by the 2030s will constitute an important step toward performing tests of GR at the smallest cosmological scales, but will be limited by control of cluster systematic effects present in galaxy-centric surveys. In the 2040s, facilities like the WST, which will combine a large FoV with a high- and low-resolution (HR/LR) MOS, have the potential to improve upon the constraints obtained with facilities in the 2030s. Given that WST will allow us to obtain spectroscopic redshifts for many LSST clusters and members, we will be in a position to exploit gravitational redshift measurements optimally.
\linebreak

\begin{figure}[ht]
\centering
\includegraphics[width=\linewidth]{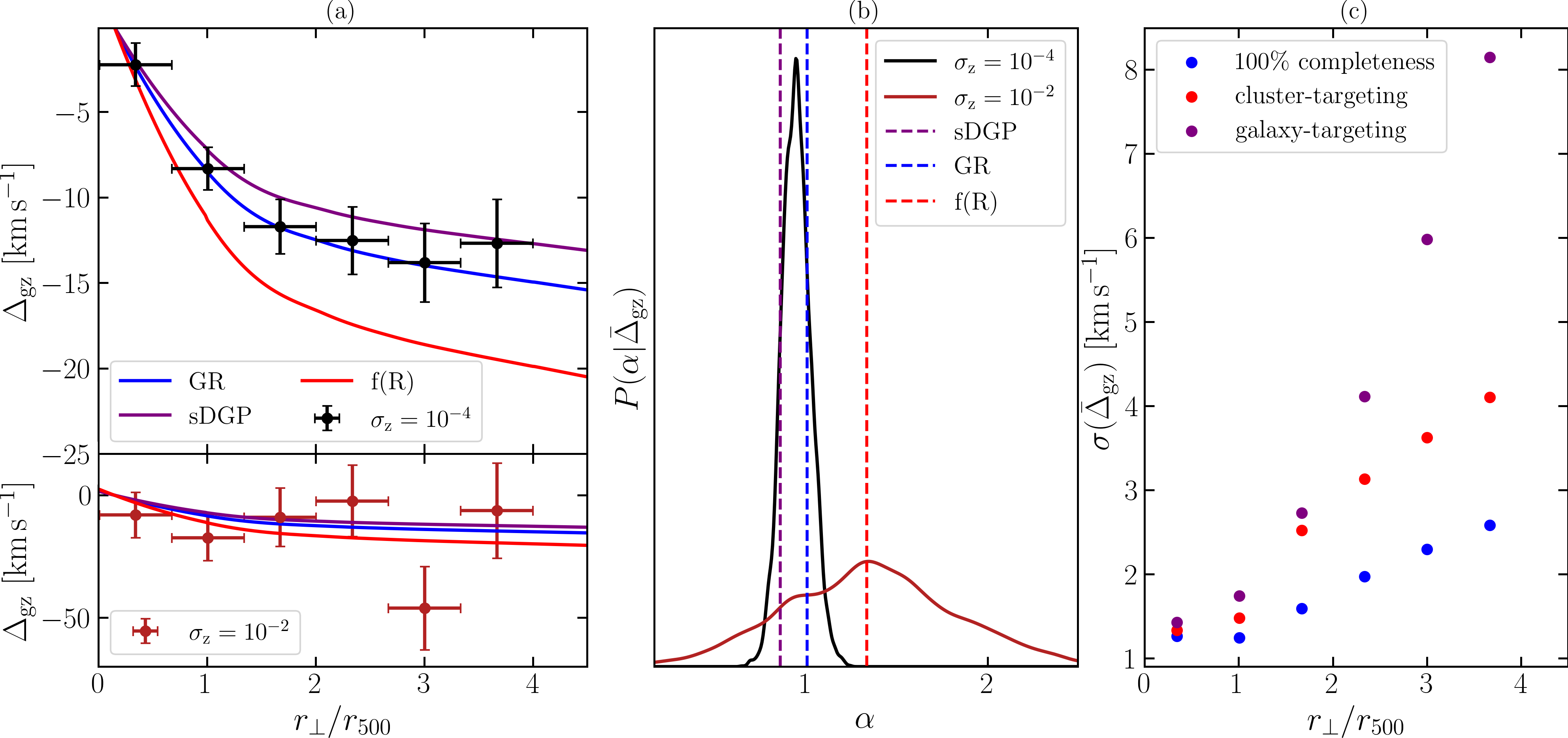}
\caption{(a) Mock gravitational redshift signal (black) as a function of projected distance from the cluster centre normalised to $r_\mathrm{500}$, compared to the binned theoretical prediction for GR (blue), sDGP (self-accelerating Dvali-Gabadadze-Porrati) gravity (purple) and f(R) gravity (red) models described in \cite{6}. In the top plot, we show a simulated data set assuming a redshift uncertainty of $\sigma_\mathrm{z}$ = $10^{-4}$, whereas in the bottom plot $\sigma_\mathrm{z}$ = $10^{-2}$ is assumed. (b) Corresponding posteriors on the modified gravity parameter, $\alpha$. The self-consistency of the pipeline has been validated to the high-S/N limit. Any fluctuations are due to the random velocity dispersion and redshift-uncertainty realisations. Our analysis assumes 560\,000 clusters out to z = 2 and 1\,400\,000 spectroscopically identified members, using a mass- and redshift-dependent halo completeness. (c) Gravitational redshift error bars as a function of projected distance from the cluster centre normalised to $r_\mathrm{500}$, for different completeness scenarios. A perfect completeness scenario is shown in blue, a galaxy-like completeness scenario in purple, and a cluster-like completeness scenario in red.}
\label{fig:forecasts}
\end{figure}

\noindent \textbf{Acknowledgements} This work was supported by STFC through Imperial College Astrophysics Consolidated Grant ST/W00 0989/1 and the FoNS Researcher Mobility Grant 2025. 

\printbibliography
\end{document}